\documentclass[final,1p,times]{elsarticle}

\usepackage{graphics}
\usepackage{graphicx}
\usepackage{epsfig}
\usepackage{amssymb}
\usepackage{amsthm}

\begin{document}
\begin{frontmatter}

\title{Measurement of the invariant mass distributions for the
$pp \to pp\eta^{\prime}$ reaction at excess energy of Q~=~16.4~MeV}

\author[1,2,3]{P.~Klaja}\ead{p.klaja@fz-juelich.de}
\author[1,2]{P.~Moskal}\ead{p.moskal@fz-juelich.de}
\author[1,2]{E.~Czerwi\'nski}
\author[1]{R.~Czy\.zykiewicz}
\author[4]{A.~Deloff}
\author[1]{D.~Gil}
\author[2]{D.~Grzonka}
\author[1]{L.~Jarczyk}
\author[1]{B.~Kamys}
\author[5]{A.~Khoukaz}
\author[1,2]{J.~Klaja}
\author[6]{K.~Nakayama}
\author[2]{W.~Oelert}
\author[2]{J.~Ritman}
\author[2]{T.~Sefzick}
\author[7]{M.~Siemaszko}
\author[1]{M.~Silarski}
\author[1]{J.~Smyrski}
\author[5]{A.~T\"aschner}
\author[2]{M.~Wolke}
\author[1]{J.~Zdebik}
\author[1,2]{M.~Zieli\'nski}
\author[7]{W.~Zipper}

\address[1]{Institute of Physics, Jagellonian University, PL-30-059 Cracow, Poland}
\address[2]{Institut f{\"u}r Kernphysik, Forschungszentrum J\"{u}lich, D-52425 J\"ulich, Germany}
\address[3]{Physikalisches Institut, Universit{\"a}t Erlangen--N{\"u}rnberg, D-91058 Erlangen, Germany}
\address[4]{Institute for Nuclear Studies, PL-00-681 Warsaw, Poland}
\address[5]{Institut f{\"u}r Kernphysik, Westf{\"a}lische Wilhelms--Universit{\"a}t, D-48149 M{\"u}nster, Germany}
\address[6]{Department of Physics, University of Georgia, GA-30602 Athens, USA}
\address[7]{Institute of Physics, University of Silesia, PL-40-007 Katowice, Poland}

\begin{abstract}
The proton-proton and proton-$\eta^{\prime}$ invariant mass
distributions have been determined for the $pp \to pp\eta^{\prime}$ reaction at an excess energy of Q = 16.4 MeV.
The measurement was carried out using the COSY-11 detector setup and the proton beam of the
cooler synchrotron COSY.
The shapes of the determined invariant mass distributions are similar to
those of the $pp \to pp\eta$ reaction and reveal an enhancement for large relative proton-proton momenta.
This result, together with the fact that the
proton-$\eta$ interaction is much stronger that the proton-$\eta^{\prime}$ interaction,
excludes the hypothesis that
the observed enhancement is caused by the interaction between the proton and the meson.
\end{abstract}

\begin{keyword}
pseudoscalar mesons \sep differential distributions \sep near threshold meson production \sep interaction
\PACS  13.60\sep 13.75.-n\sep 14.40.-n\sep 25.40.-h
\end{keyword}

\end{frontmatter}


\section{Introduction}
\label{sec:intro}
The understanding of the meson-nucleon interaction as well as studies of meson structure
and production mechanisms constitute one of the basic issues of the contemporary
hadron physics.
The $\eta$ and $\eta^{\prime}$ mesons constitute a mixture of the SU(3) singlet and octet states
with almost the same relative contributions of various quark flavours.
Nevertheless, they have unexpectedly different properties, e. g. as regards the mass \cite{pdg},
branching ratios \cite{jossop, branden} or production cross sections \cite{prc69}.
These differences indicate that also the interaction of
$\eta$ and $\eta^{\prime}$ mesons with nucleons may be different.\\
Up to now, the proton-$\eta$ interaction was studied 
intensively
but
still rather large range of values of the scattering length is reported
depending on the analysis method \cite{wycech}.
The proton-$\eta^{\prime}$ interaction  is much less known.
It is only qualitatively estimated (based on the $pp\to pp\eta^{\prime}$
excitation function)  to be much weaker than for the proton-$\eta$ system~\cite{swave}.
In principle,  studies of $pp\to pp\, meson$ reactions permit
information about the proton-meson interaction to be gained
not only from the shape of the excitation function but also from
differential distributions of proton-proton and proton-meson invariant masses.
Therefore, in order to investigate the proton-$\eta$ interaction the COSY-11
collaboration performed a measurement~\cite{prc69} of the proton-$\eta$
and proton-proton invariant mass distributions close to
the threshold at Q = 15.5 MeV,
where the outgoing particles possess small relative velocities.
Indeed a large enhancement
in the region of small proton-$\eta$ and large proton-proton relative momenta
was observed\footnote{
The same enhancement was also seen in independent measurements
by the COSY-TOF group~\cite{tof41}.}.
However, the observed effect cannot be univocally assigned
to the influence of the
proton-$\eta$ interaction in the final state~\cite{fix,fix2},
since it
can also  be explained
by the admixture of higher partial waves
in the proton-proton system \cite{kanzo}, or by the
energy dependence of the production amplitude~\cite{deloff,ceci}.

The endeavor to explain the origin of the observed enhancement
motivated the measurement of the proton-proton and proton-$\eta^{\prime}$
invariant mass distributions for the $pp \to pp\eta^{\prime}$ reaction
presented in this article.

If the enhancement observed in the proton-proton
invariant mass distributions for the $pp \to pp\eta$ reaction would be due to the proton--$\eta$ interaction
then it is expected to be significantly lower for the $pp \to pp\eta^{\prime}$ reaction
since the proton-$\eta$ interaction is stronger than  the proton-$\eta^{\prime}$~\cite{swave}.

In order to make a model independent comparison of the spectra
in the $pp\eta$ and $pp\eta^{\prime}$ systems we performed a measurement
of the $pp \to pp\eta^{\prime}$ reaction,
nominally at the same excess energy as previously measured
for the $pp \to pp\eta$ reaction.
Invariant mass spectra determined at the same excess energies allow for the comparison
without a need for a correction of kinematical factors in the outgoing system.

It is important to stress that the invariant mass distributions for the $pp \to pp\eta^{\prime}$ reaction
have not been measured so far.
This is because the total cross section for the $\eta^{\prime}$ meson
production in hadron collisions is by more than a factor of 30 smaller than the one for the $\eta$
meson at the same excess energy and additionally the total cross section
for the production of the multi-pion background grows by three orders of magnitude when
the beam energy increases from the $\eta$ to the $\eta^{\prime}$ production
threshold~\cite{marcin}.
A determination of the invariant mass spectra reported in this letter was made possible due
to stochastic cooling of the proton beam of the COSY synchrotron \cite{ring1, cooling, cooling1, cooling2} and the
good momentum resolution ($\sigma$ = 4 MeV/c) \cite{prc69}
achieved with the
COSY-11 detector setup designed especially for measurements near the kinematical threshold.
Sufficient statistics were collected to allow the background to be subtracted from the invariant mass distributions.

Detailed description of the COSY-11 method  used for the
$pp \to pp\eta^{\prime}$ reaction measurements
can be found in \cite{pm80, b474, khoukaz}, therefore, here
we concentrate on the background subtraction procedure crucial for the
determination of invariant mass distributions.

\section{Experiment and data analysis}
\label{sec:exp}
\begin{figure}[H]
\includegraphics[width=.3\textheight]{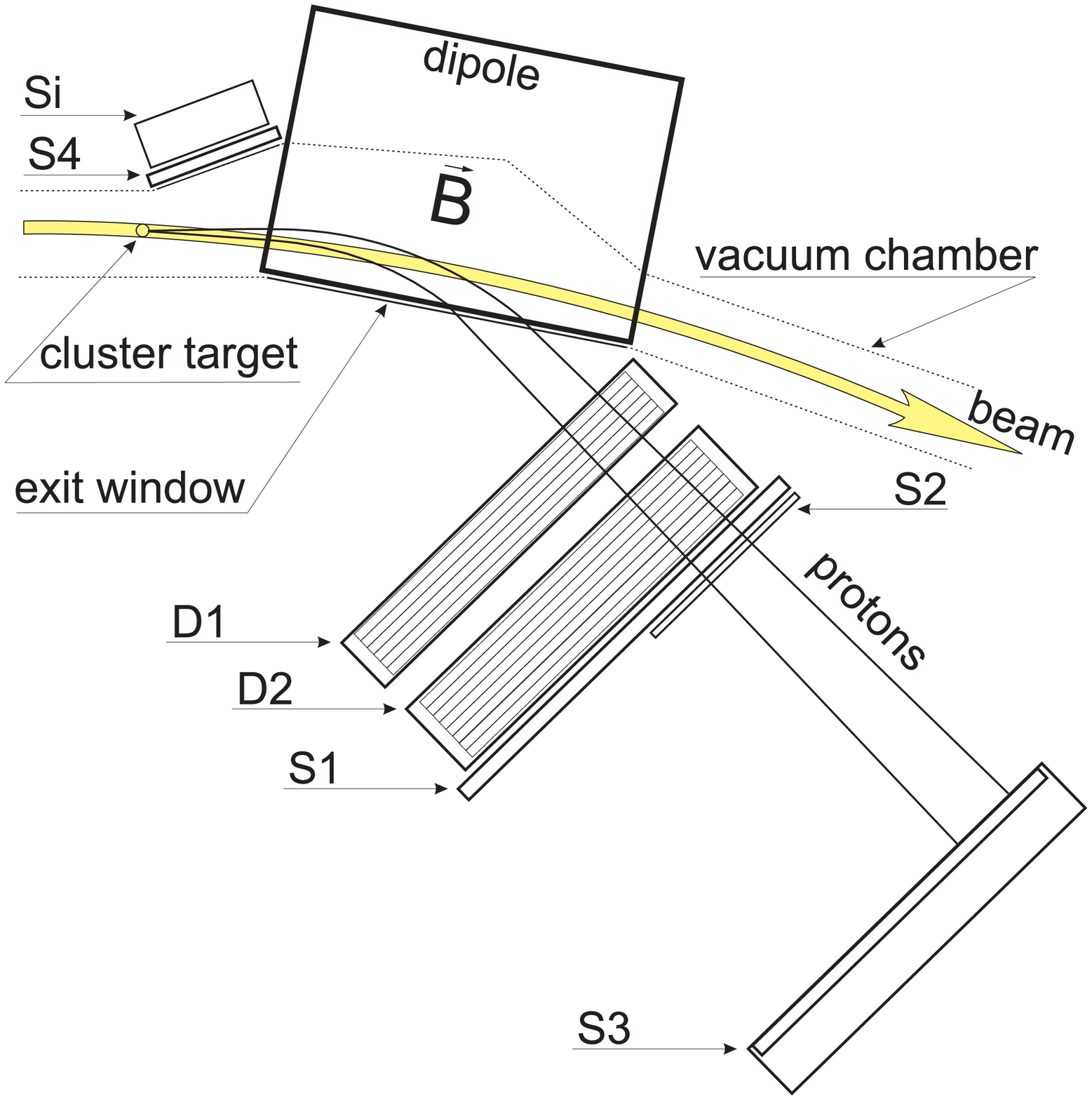}
\caption{
Schematic view of the COSY-11 detector facility \cite{brauksiepe}. 
Protons originating from the $pp \to ppX$ reaction are bent in the dipole magnetic field,
and leave the vacuum chamber through the exit window. Afterwards they are
detected in the two drift chambers D1 and D2, in the scintillator hodoscopes S1 and S2, and
in the scintillator wall S3. The scintillation detector S4 and the silicon pad detector Si are
used in coincidence with the D1, D2 and S1
detectors for the registration of the elastically scattered protons.}
\label{fig:setup}
\end{figure}

The measurement of the $pp \to pp\eta^{\prime}$ reaction was conducted using the
cooler synchrotron COSY \cite{ring1} and the COSY-11 detector
setup~\cite{brauksiepe, dombrowski, klajac11}
at the proton beam momentum of $p_{B}$~=~3.260~MeV/c, which corresponds to
an excess energy of Q~=~16.4~MeV. It was based on the registration
of the two outgoing protons and reconstruction of their momenta. The $\eta^{\prime}$ meson
is identified using the missing mass technique.\\ 
Figure \ref{fig:setup} illustrates a schematic view of the COSY-11 apparatus with a topology of 
the $pp \to ppX$ reaction. Two outgoing
protons possessing smaller momenta than the beam momentum are bent in the dipole magnetic field
towards the detector system leaving the vacuum chamber through the exit window. Afterwards, they are
detected in the two drift chambers, D1 and D2, in the scintillator hodoscopes S1 and S2, and
in the scintillator wall S3.
The target\footnote{The $H_{2}$ cluster target specifications are described in references \cite{dombrowski, target1}.}
used during the experiment, was realized as a beam of $H_{2}$ molecules grouped inside clusters of up to about $10^{6}$~atoms.
The average density of the target was around $5 \cdot 10^{13}$ atoms/cm$^{2}$ \cite{target1}.
It was installed in front of the dipole magnet as it can be seen schematically in Figure \ref{fig:setup}.\\

In order to determine the absolute values of the differential cross sections, the
time integrated luminosity ($L$) has been established by
the concurrent measurement of the angular distribution of the elastically
scattered protons \cite{nim}.
The extracted value of the integrated luminosity 
amounts to $L~=~(5.859 \pm 0.055)~pb^{-1}$ \cite{pk_phd}.

\begin{figure}[H]
\includegraphics[width=.3\textheight]{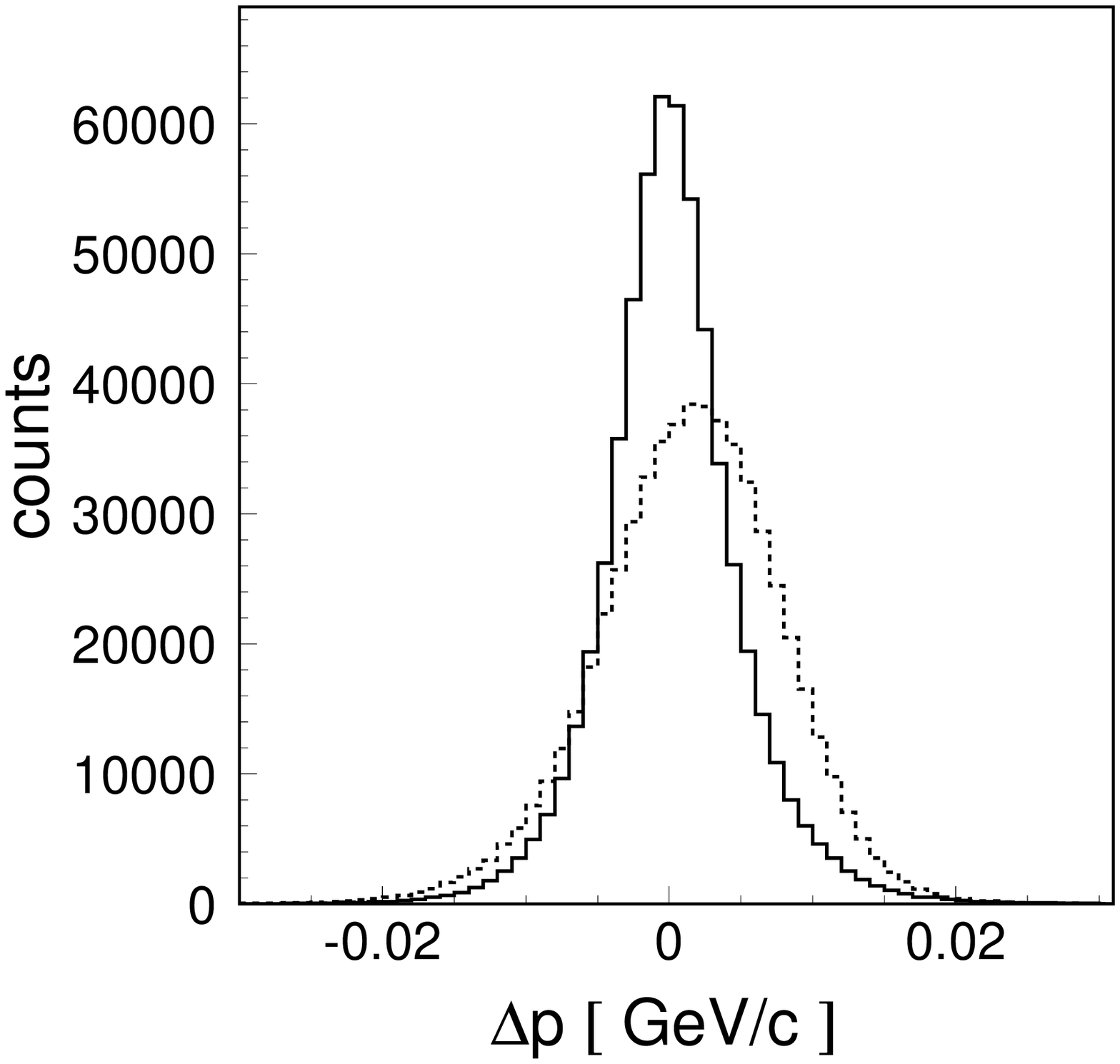}
\caption{Spectrum of the difference between the simulated and reconstructed proton momentum. The dashed
                      line denotes the spectrum before kinematical fit and the solid line corresponds
                      to the situation after the fitting procedure.}
\label{fig:delta_p}
\end{figure}

\begin{figure}[H]
 \includegraphics[height=.3\textheight]{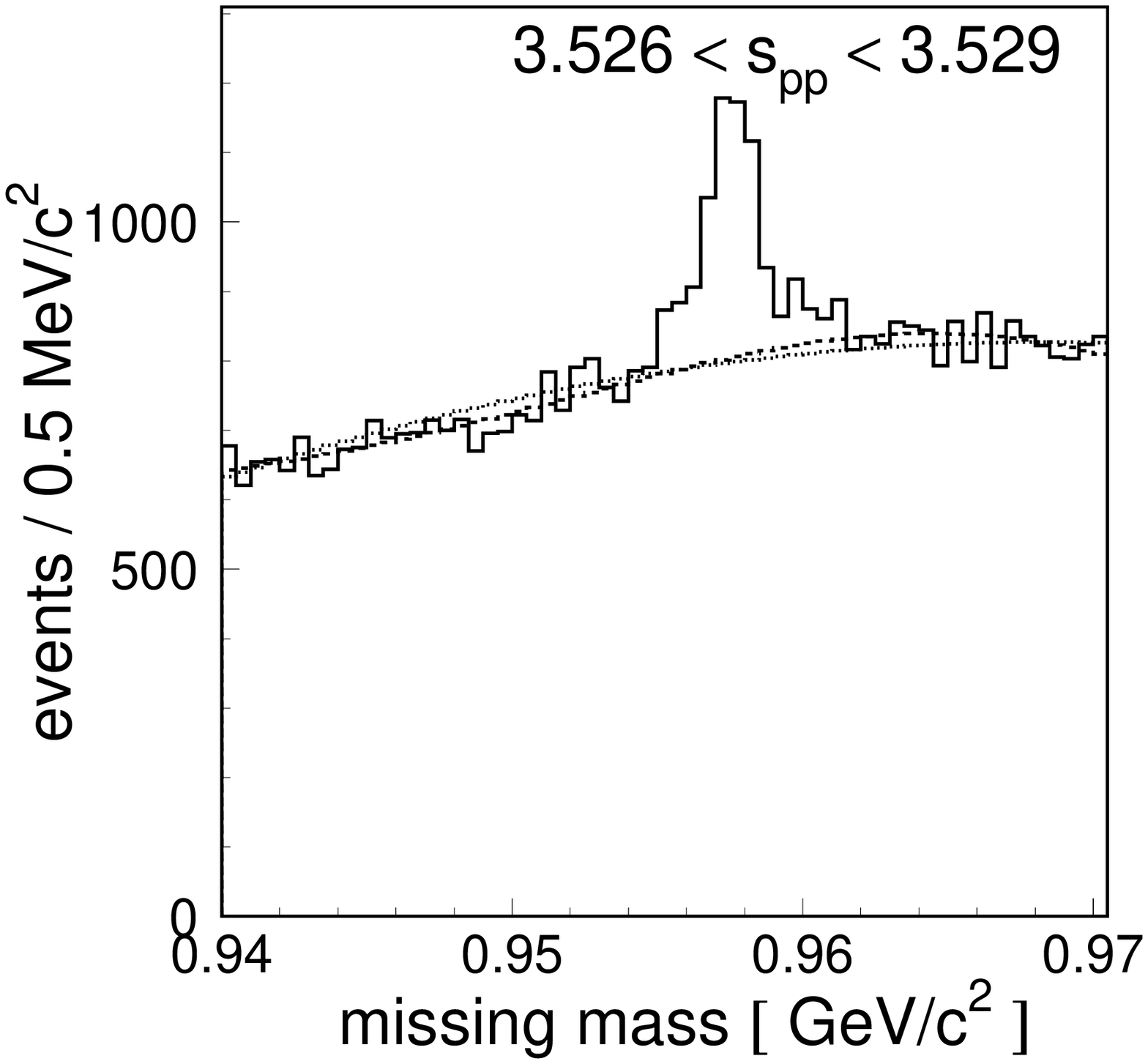}
  \includegraphics[height=.3\textheight]{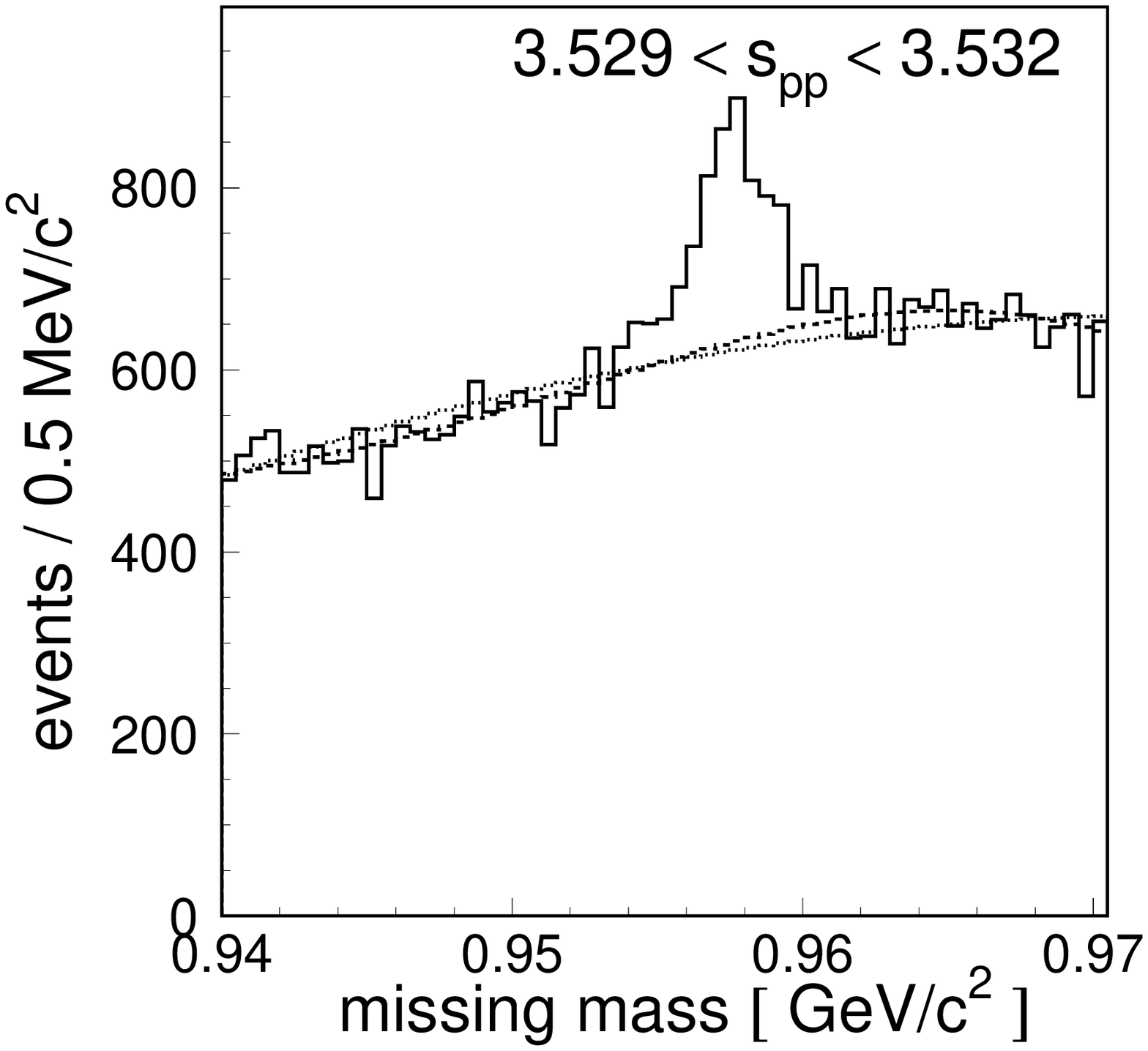}
          \caption{Examples of experimental missing mass spectra for two intervals of
          $s_{pp}$ as indicated in the figure. The dotted lines indicate second order polynomials and
          the dashed lines show the sum of two Gaussian distributions. }
          \label{fig:gauss_pol}
\end{figure}

In order to search for small effects like proton-meson interaction
it is of importance to account for smearing of the measured distributions
due to the finite resolution of the detector system, which may
alter the shape of the spectrum especially close to the kinematical limit.
Therefore,
as has been done previously in the analysis of the $pp \to pp\eta$ reaction \cite{prc69}, a kinematical fitting 
of the data has been performed \cite{pk_phd} in order to improve resolution. 
To this end, the momenta of the protons were varied demanding that the missing mass of the unregistered
particle equals the known mass of the $\eta^{\prime}$ meson exactly. Furthermore, as a result of the fit, only those proton 
momentum vectors which
were closest to the vectors determined from the experiment have 
been chosen. 
The inverse of the covariance matrix was used as a metric 
for the distance calculation. The kinematical fit improves the resolution by a factor of 1.5,
as can be seen in Figure \ref{fig:delta_p}.
After the kinematical fit  each event 
can be characterized by both: 
experimentally determined
momentum vectors and kinematically fitted momenta.
In the subsequent analysis the fitted momenta were used to 
group events into $s_{pp}$ and $s_{p\eta^{\prime}}$ intervals
and then in order to subtract the background,
for each group separately, a missing mass spectrum was determined
from experimental momentum vectors. The available range of $s_{pp}$ and $s_{p\eta^{\prime}}$
was divided into 22 bins. The width of the bins (0.003~GeV$^{2}$/c$^{4}$) was chosen as a compromise between statistics and the
experimental resolution.
Then, for each bin, the missing mass spectrum was reconstructed
and the number of the $pp \to pp\eta^{\prime}$ events was calculated separately for each interval of $s_{pp}$, $s_{p\eta^{\prime}}$.\\
In Figure \ref{fig:gauss_pol}, examples of missing mass spectra 
for two 
intervals of
the invariant proton-proton mass are presented.
A clear signal from the $\eta^{\prime}$ meson production is seen on top
of a continuous spectrum
from the multi pion background. A~smooth behaviour of the background allows one to
interpolate its shape under the $\eta^{\prime}$ peak with polynomial functions that match the data
below and above the peak. The smooth behaviour of the multi-pion background in the peak region was verified by Monte Carlo 
simulations \cite{pk_phd}.
In both panels of Figure \ref{fig:gauss_pol} the dotted lines correspond to the
result of the fit of the second order polynomial. An equally good approximation of the
background was also achieved by a fit of the sum of two Gaussian distributions~\cite{pk_phd}.
The parameterizations were performed in
the entire range of missing mass outside of the $pp \to pp\eta^{\prime}$ signal,
and obviously reproduce the background very well.

\begin{figure}[H]
  \includegraphics[height=.3\textheight]{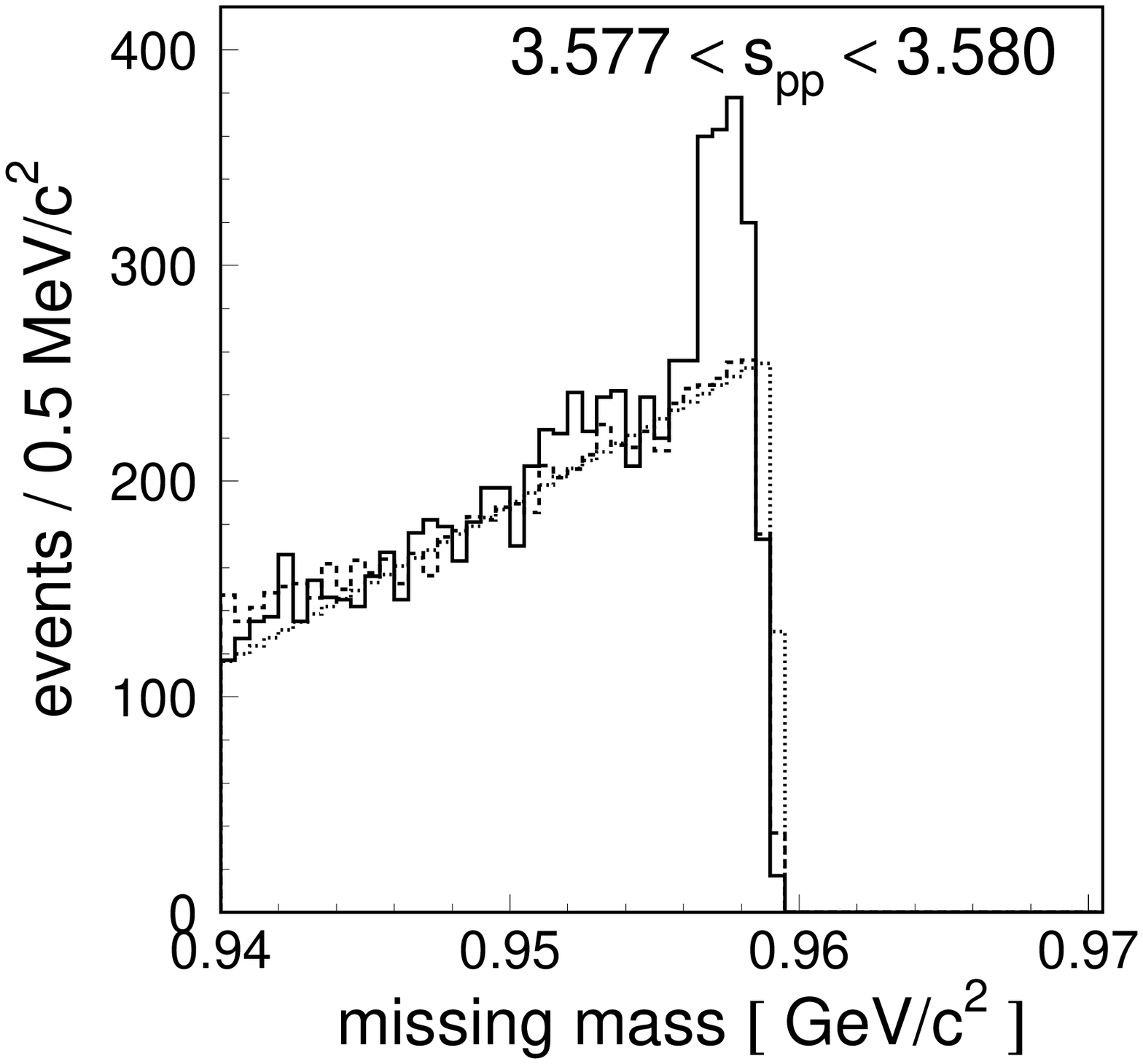}
  \includegraphics[height=.3\textheight]{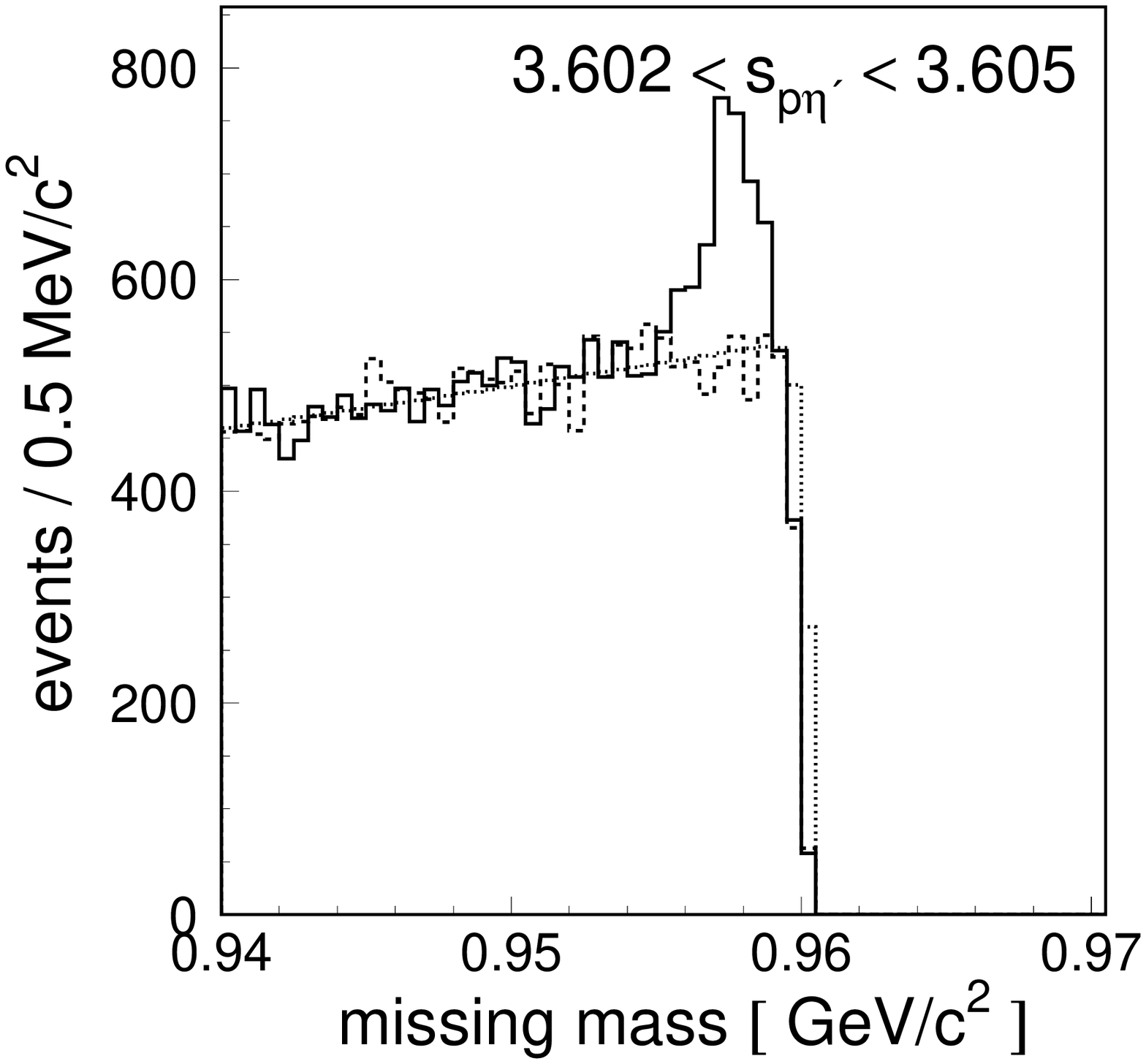}
          \caption{Examples of measured missing mass spectra for the $pp \to ppX$
          reaction (solid lines) with superimposed Monte-Carlo simulations of multi-pionic background (dashed lines).
          The missing mass spectra are presented for squared invariant proton-proton mass $s_{pp} \in [3.577;~3.580]~ $GeV$^{2}$/c$^{4}$
           (left) and
          squared invariant proton-$\eta^{\prime}$ mass $s_{p\eta^{\prime}} \in [3.602;~3.605]~ $GeV$^{2}$/c$^{4}$ (right).
          The dotted lines in both panels correspond to the fit of  Equation \ref{equ:fermi}.}
          \label{fig:exp_mc}
\end{figure}

\begin{figure}[H]
  \includegraphics[height=.3\textheight]{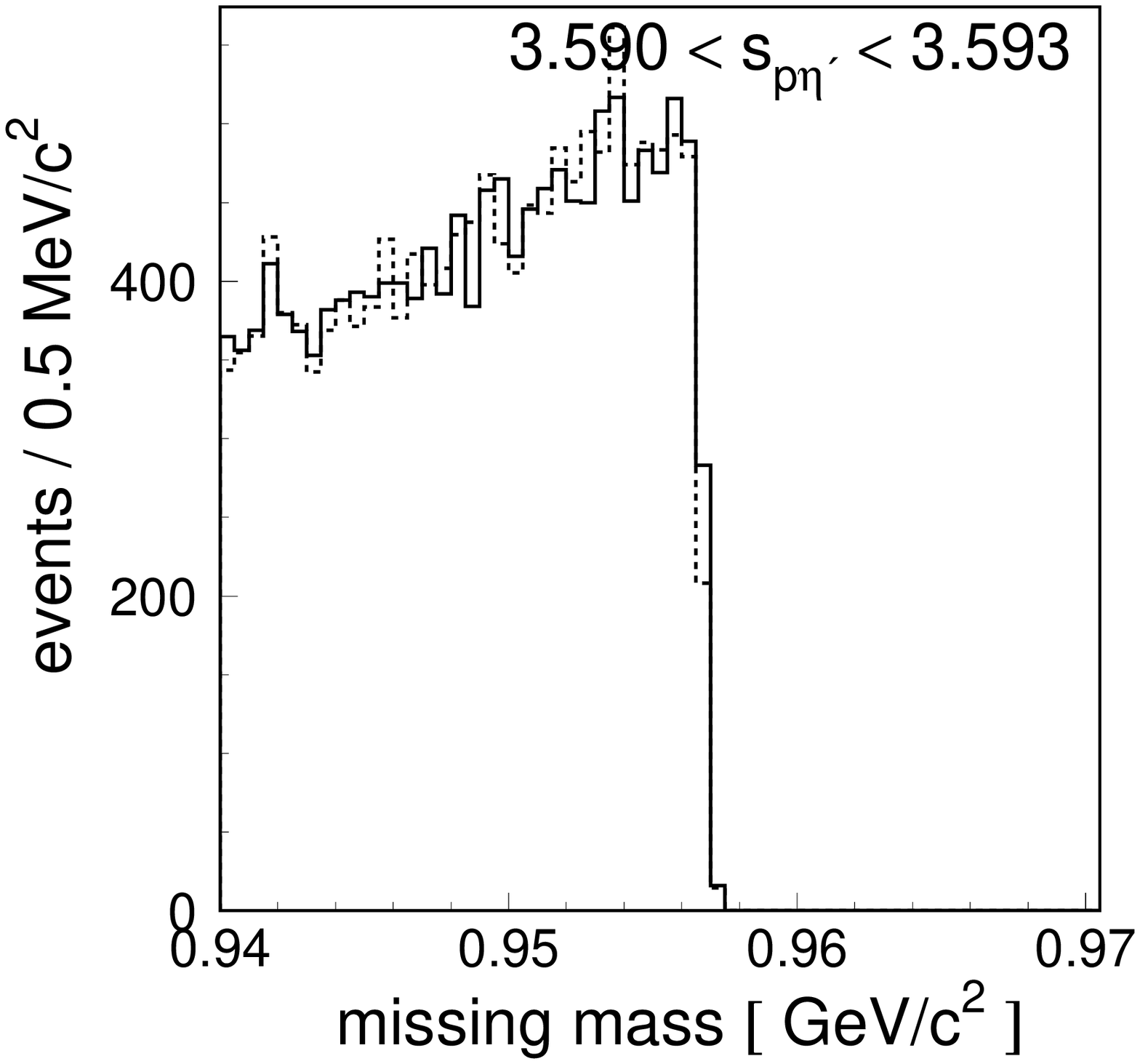}
  \includegraphics[height=.3\textheight]{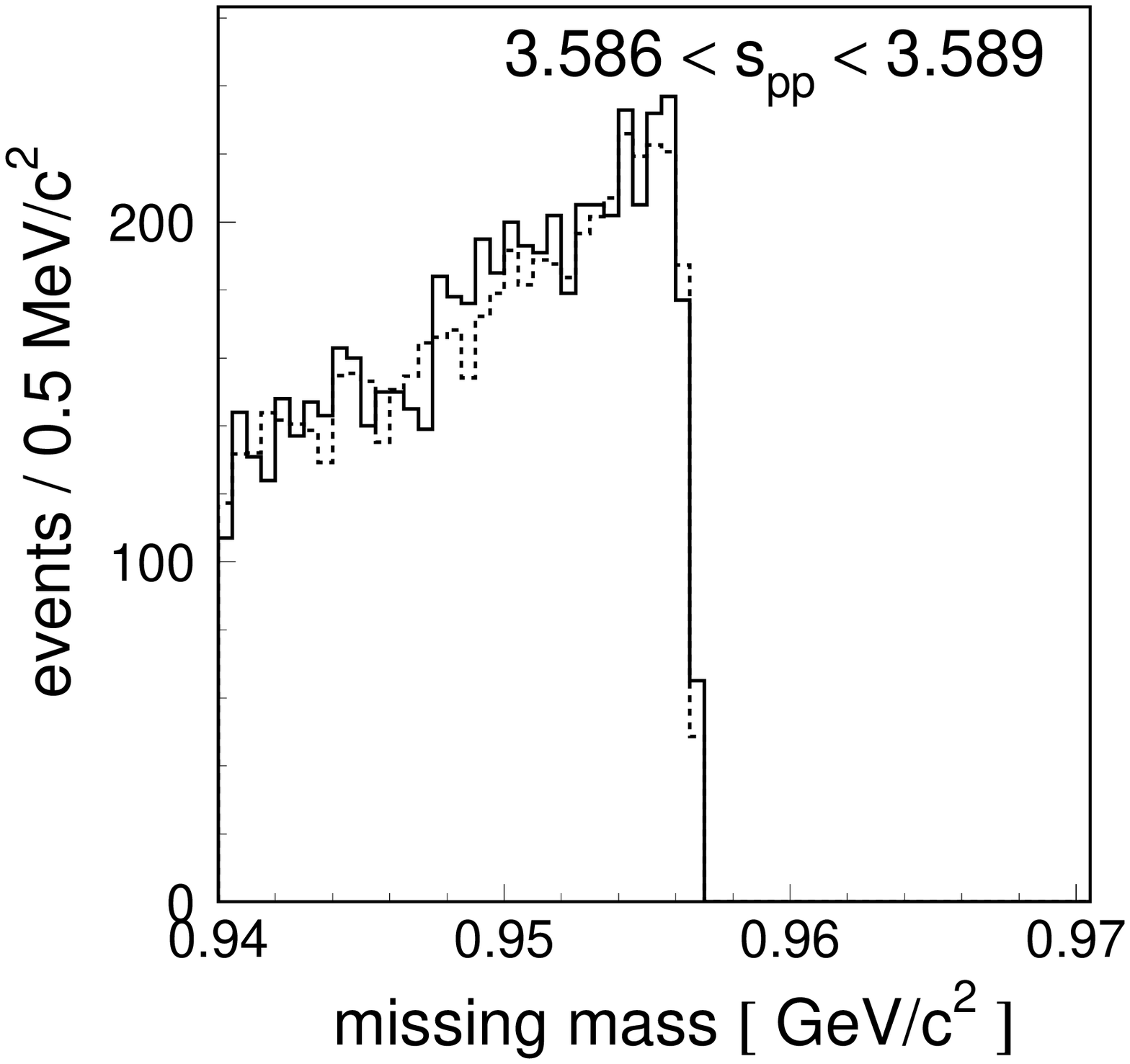}
          \caption{Missing mass spectra for low values of $s_{p\eta^{\prime}} \in [3.590;~3.593]~$GeV$^{2}$/c$^{4}$ (left) and high
          values of $s_{pp} \in [3.586;~3.589]~$GeV$^{2}$/c$^{4}$~(right). The dashed
          lines correspond to a linear sum of the
          simulated $pp \to pp2\pi\eta$, $pp \to pp3\pi$ and $pp \to pp4\pi$
          reactions fitted to experimental distributions (solid lines) using only the magnitudes as free parameters.}
          \label{fig:exp_pions}
\end{figure}

The situation is more complicated for these missing mass spectra when the signal is
close to the kinematical limit (see Fig. \ref{fig:exp_mc}). In this case the shape of the background
on the right side of the peak cannot be easily predicted.
Such spectra are obtained for kinematical regions of higher
squared invariant proton-proton masses
 and relatively low squared invariant proton-$\eta^{\prime}$ masses. In order to describe the shape of the background
in these regions, the $pp \to pp2\pi\eta$, $pp \to pp3\pi$ and $pp \to pp4\pi$ reactions\footnote{These 
 reaction channels
were chosen as a representation of the possible multi-pion production background, 
since contribution from the $pp \to pp5\pi$,$6\pi$,$7\pi$
reactions can be neglected \cite{pm80}
and the missing mass spectrum from $2\pi$ has a similar
shape as those for $3\pi$ and $4\pi$ in the relevant kinematical region.}
have been generated and the simulated events were analysed in the same way as
for the experimental data. The result of these simulations (dashed lines) is compared to the experimental data in 
Figure~\ref{fig:exp_mc}.
The simulations of the different reaction channels were performed with a distribution based on an equal population of phase space
including the proton-proton final state interaction~\cite{swave, jpg}.
A linear sum of the simulated missing mass spectra was matched to the data with the relative magnitudes as the only free parameters.
In both examples, simulations are in a good agreement with
the experimental background distributions below the $\eta^{\prime}$ peak. Moreover the behaviour of the simulated background
matches the kinematical limit of the missing mass distributions.\\
For the dynamics of the pion production, it had been assumed that pions are produced uniformly over the available phase space.
As described in reference \cite{jpg} the shape of the missing mass
spectrum does not change significantly at the edge of the kinematical limit if one assumes resonant
or direct pion production.\\
In order to increase the confidence in the estimation of the background behaviour near the kinematical boundary
and to estimate the systematic uncertainties due to the choice of the background parameterizations,
these distributions were described in an independent way with a second order polynomial divided by the Fermi
function for the description of the rapid slope at the end of the distributions.
To this end, the following formula was applied:
\begin{equation}
  F(mm,a,b,c,d,g)~=~(a~+~b \cdot mm~+~c \cdot mm^{2})/(1~+~e^{(mm~-~d)/g}),
\label{equ:fermi}
\end{equation}
where $a$, $b$, $c$, $d$ and $g$ are free parameters.\\
The results are presented in Figure \ref{fig:exp_mc} as dotted lines.
It is seen that under the $\eta^{\prime}$ peak the result of Equation \ref{equ:fermi}
agrees well with the background determined
from the simulations and that both reproduce the shape of the slope quite well.\\
A further check of the background was performed by extracting 
the missing mass distributions
for regions of the squared invariant masses of proton-proton and proton-meson
where the $\eta^{\prime}$ is not produced.
The resulting missing mass distributions are shown in Figure \ref{fig:exp_pions} and
represent the regions of low squared invariant masses of the proton-meson subsystem (left)
and high squared invariant masses of the proton-proton subsystem (right).
For such values of $s_{pp}$ or $s_{p\eta^{\prime}}$ the production of the $\eta^{\prime}$ meson is not kinematically allowed
because $s_{p\eta^{\prime}} < (m_{p} + m_{\eta^{\prime}})^{2}$ and $s_{pp}$ is too large leaving insufficient energy to create the
$\eta^{\prime}$ meson. The simulations reproduce the background very well as one can see in Figure \ref{fig:exp_pions}.\\
The main contribution to the systematic uncertainty of the determined
differential cross sections  comes from
the uncertainty of the estimation of the yield of the $\eta^{\prime}$ events which, in turn, is due to the assumption of 
the shape of the background.
In order to estimate these errors, the numbers of background events
extracted under the two different assumptions were compared.
For the missing mass spectra with the signal far from the kinematical limit, 
the background determined by Gaussian distributions was compared to the background estimated by a second order polynomial.
For the spectra close to the kinematical limit, the background determination by
Monte-Carlo simulations was compared to a second order polynomial divided by the Fermi distribution.
The uncertainty of the background estimation constitutes the main contribution to the systematic error
of the differential cross sections. 
The differences in the number of background events obtained by applying the above described different
fit procedures are below $3\%$ of the background value,
which corresponds to about $20\%$ of the signal, as can be seen for instance in Figure \ref{fig:exp_mc}.

\begin{figure}[H]
\includegraphics[width=0.3\textheight]{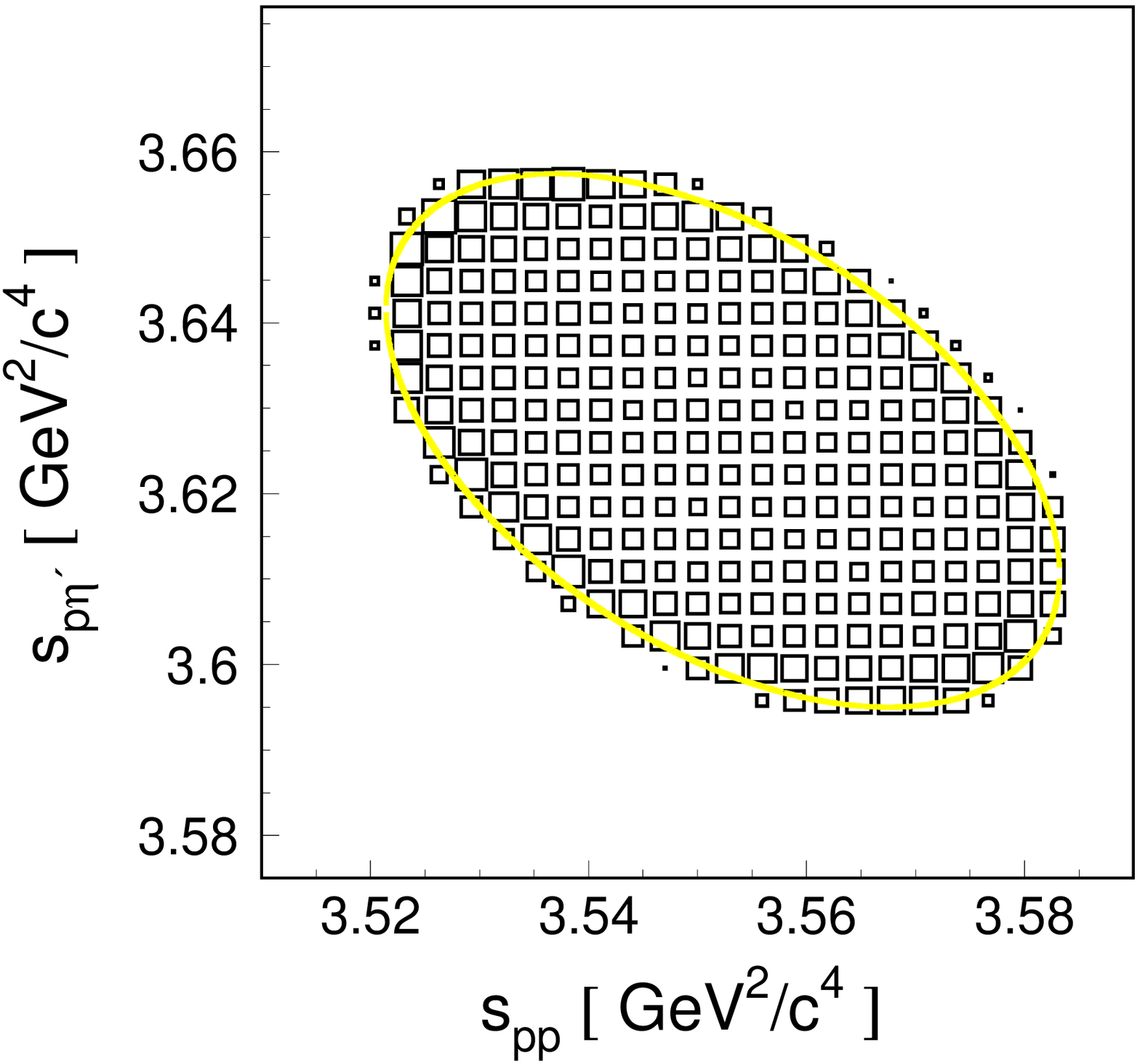}
\caption{COSY-11 detection acceptance as a function of $s_{pp}$ and $s_{p\eta^{\prime}}$ 
squared invariant masses. Dalitz 
plot distribution was reconstructed from the events accepted by the COSY-11 detector setup.}
\label{fig:dalitz}
\end{figure}

Finally, the measured distributions were corrected for the acceptance according to the method described elsewhere \cite{prc69}.
Here, it is important to stress that
at the beam momentum of $p_{B}$ = 3.260 GeV/c the COSY-11 acceptance
for the $pp\to pp\eta^{\prime}$ reaction is finite over the entire
area of the $s_{pp}$~vs~$s_{p\eta^{\prime}}$ Dalitz plot.
This is shown in Figure \ref{fig:dalitz}, where one can see that the full phase
space for the $pp \to pp\eta^{\prime}$ reaction is covered.\\

\section{Results}
\label{sec:results}
\begin{figure}[H]
 \includegraphics[height=.275\textheight]{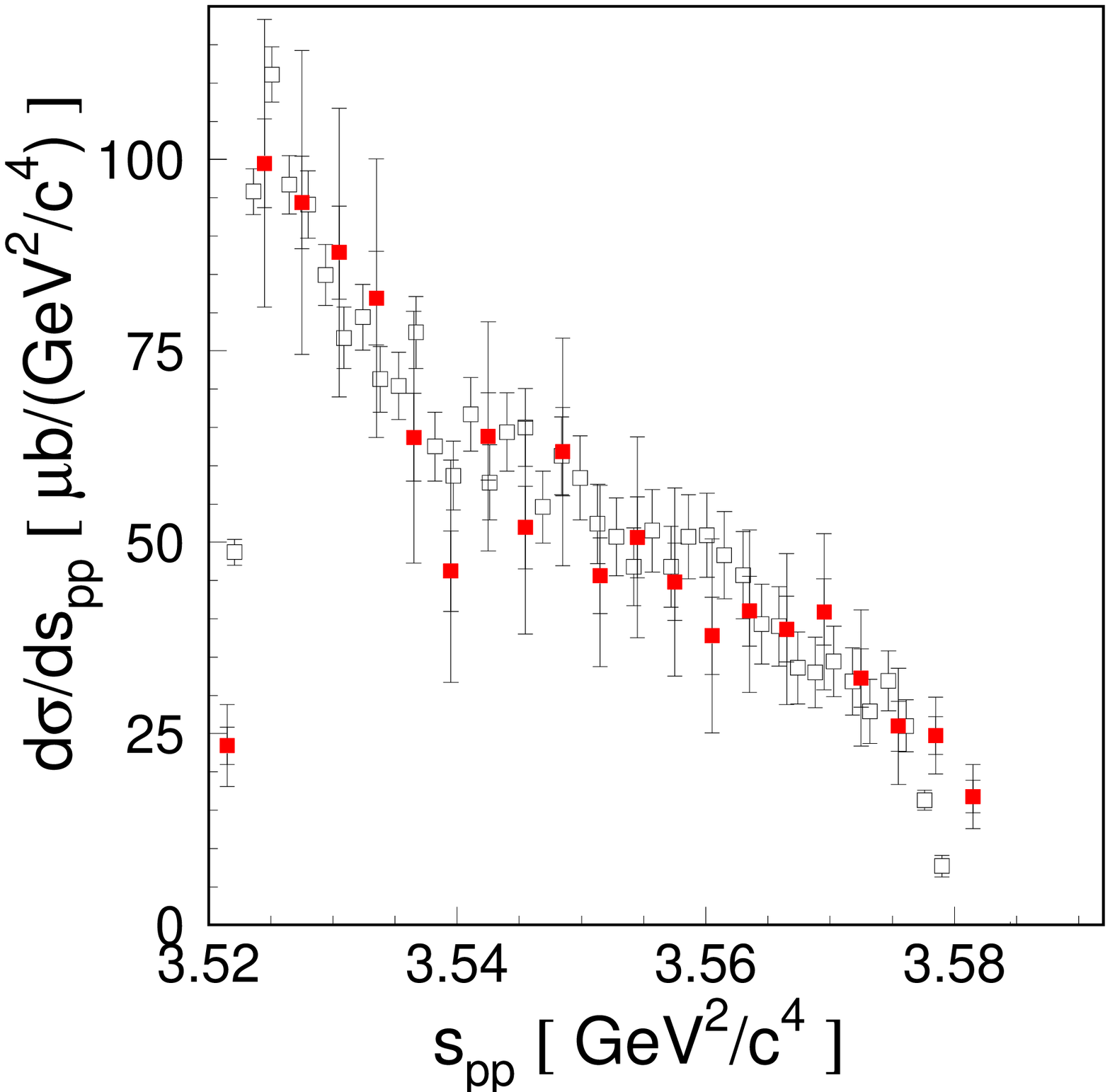}
  \includegraphics[height=.275\textheight]{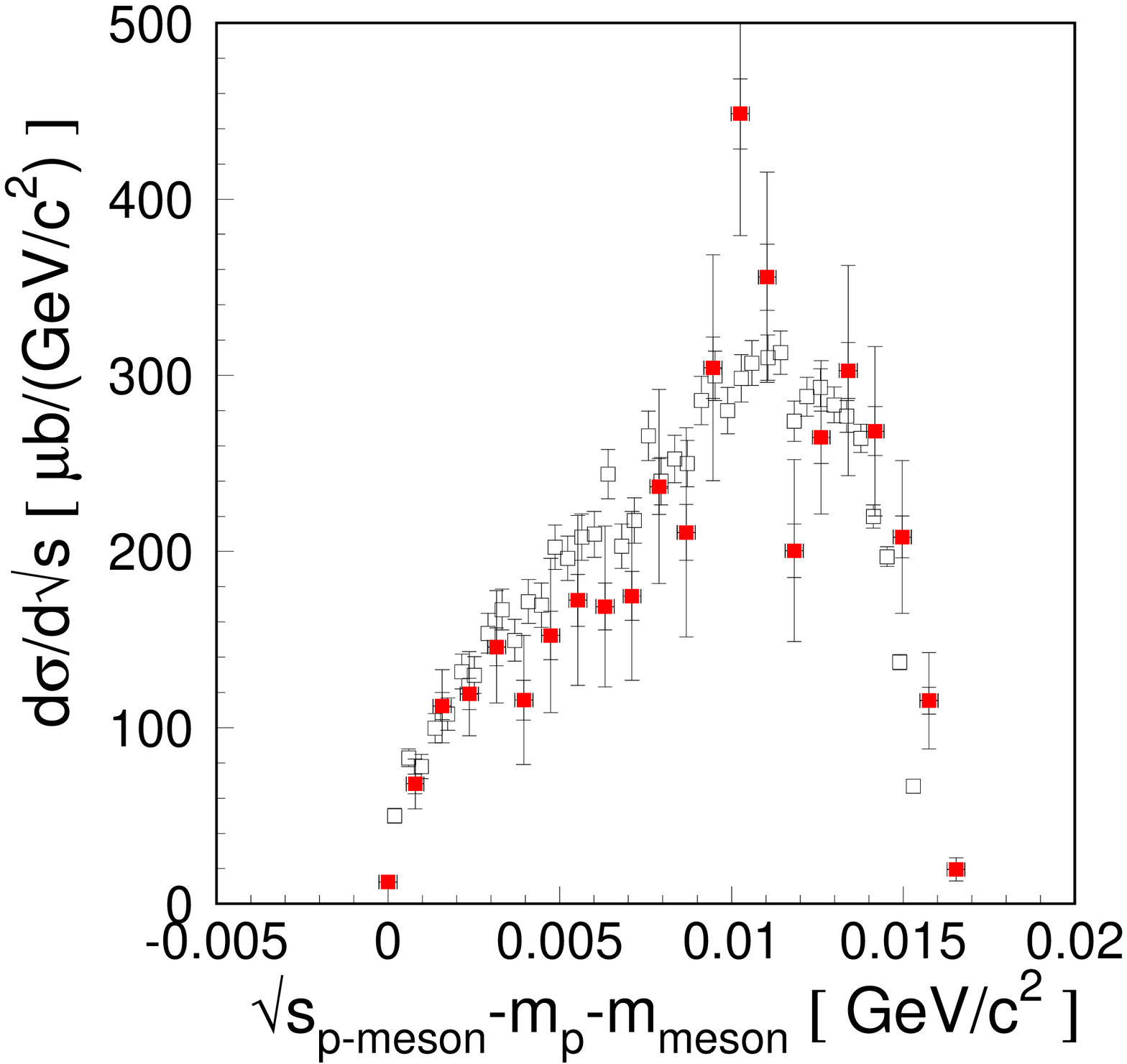}
        \caption{Comparison of the proton-proton invariant mass squared distributions ($s_{pp}$) (left)
        and of proton-meson kinetic energy ($\sqrt{s_{p-meson}}~-~m_{p}~-~m_{meson}$) (right).
        The distributions for the $pp \to pp\eta^{\prime}$
        reaction (filled squares)
        were normalized to the same total cross section as the $pp \to pp\eta$ reaction (open squares).
	    Statistical and systematic errors were separated by horizontal dashes.}
        \label{fig:comparison}
\end{figure}

\begin{figure}[H]
 \includegraphics[height=.3\textheight]{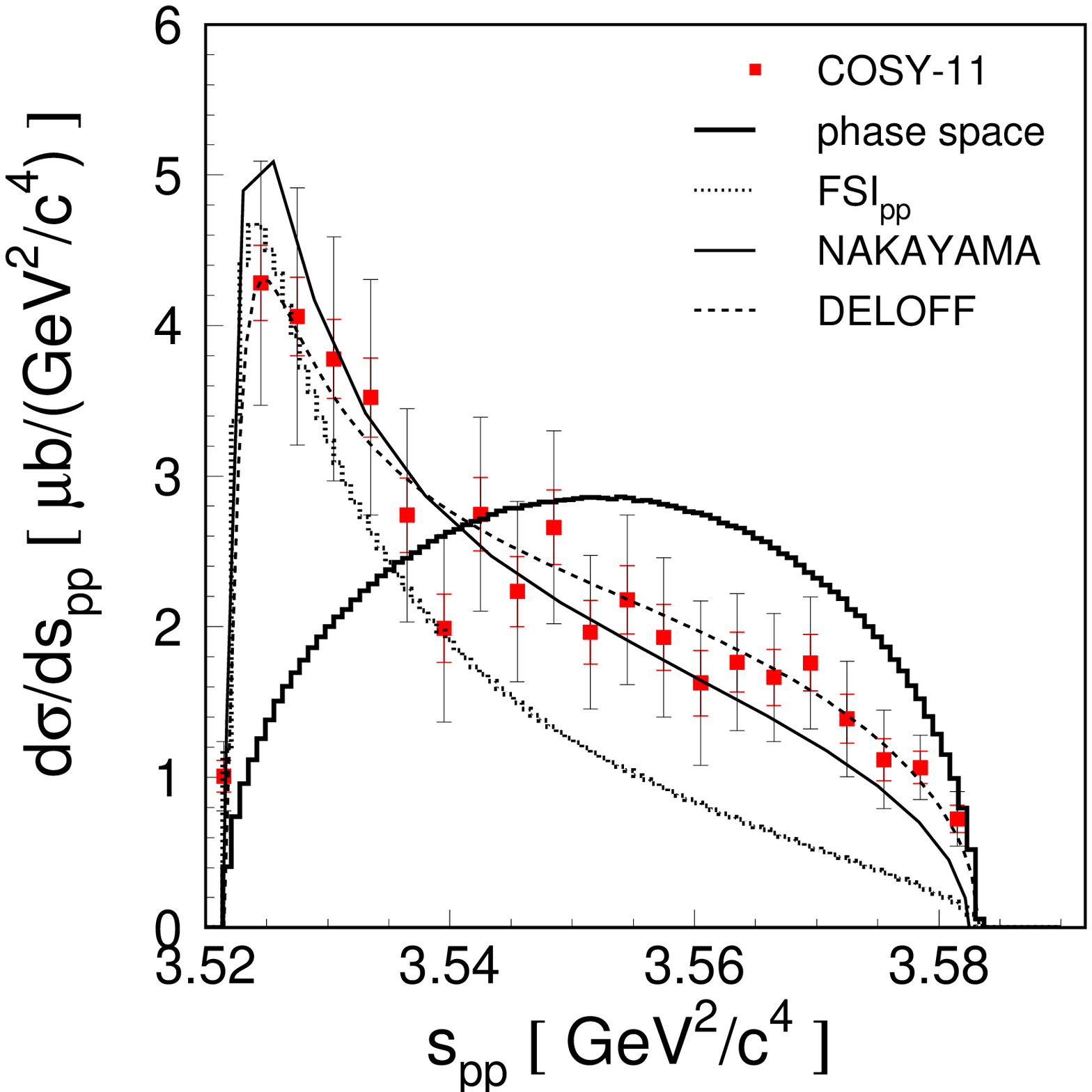}
  \includegraphics[height=.3\textheight]{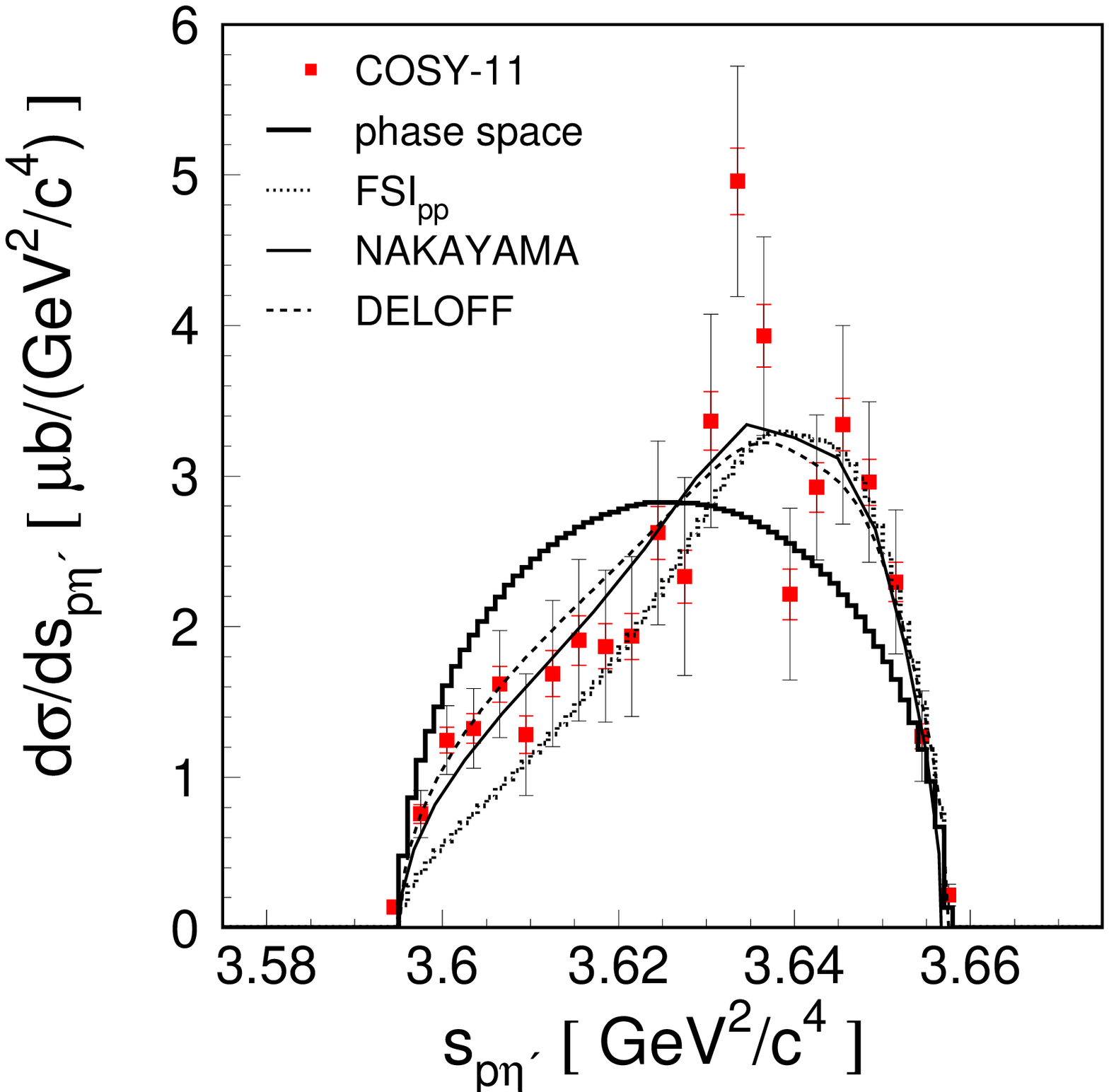}
        \caption{Distributions of the squared proton-proton ($s_{pp}$)
        and proton-$\eta^{\prime}$ ($s_{p\eta^{\prime}}$) invariant masses,
        for the $pp \to pp\eta^{\prime}$ reaction at the excess energy of Q = 16.4 MeV.
        The experimental data (closed squares) are
        compared to the expectation under
 the assumption of a homogeneously populated phase space
        (thick solid lines)
and the integrals of the phase space weighted
        by the proton-proton scattering amplitude - FSI$_{pp}$ (dotted histograms).
	The solid and dashed lines correspond to calculations when taking into account contributions from higher partial waves
	and allowing for a linear energy dependence of the $^{3}P_{0} \to ^{1}\!\!S_{0}s$ partial wave amplitude, respectively.
         }
        \label{fig:norm_pp_peta}
\end{figure}

A comparison of the $s_{pp}$
and $\sqrt{s_{p-meson}}$ 
distributions 
between the $pp \to pp\eta^{\prime}$ and the $pp \to pp\eta$ reactions is presented in Figure \ref{fig:comparison}.
For the proton-meson system the comparison was performed for the
kinetic energy ($\sqrt{s_{p-meson}}~-~m_{p}~-~m_{meson}$)
and not as a function of $s_{p-meson}$ because the range of the $s_{p\eta}$ and $s_{p\eta^{\prime}}$ are different
due to the different masses of the $\eta$ and $\eta^{\prime}$ mesons.
But the range of ($\sqrt{s_{p-meson}}~-~m_{p}~-~m_{meson}$) is the same since
the measurements
for the $\eta$ and $\eta^{\prime}$ production were performed at about
the same excess energy\footnote{The nominal beam energy
corresponds to an excess energy of Q = 15.5 MeV, consistent with the data collected for the
$pp \to pp\eta$ reaction. However, the real value was determined to be 16.4 MeV~\cite{pk_phd}. This difference is well within
the precision of the absolute beam momentum adjustment of the COSY synchrotron amounting to 
about $\delta p / p$~=~10$^{-3}$ \cite{cooling}
which corresponds to $\sim$1 MeV uncertainty in Q.}.\\
In both panels it is seen that the shape of the distribution for the $pp \to pp\eta$
measurement (open squares) is in agreement with that for the $pp \to pp\eta^{\prime}$ reaction (closed squares) within the error bars,
thus, showing the same enhancement
in the region of large proton-proton invariant mass.
Therefore, if the $\eta^{\prime}$-proton interaction
is indeed much smaller than the $\eta$-proton
as inferred from the excitation function \cite{prc69,swave}, then, 
one would have observed a significantly smaller enhancement in the case
of the $\eta^{\prime}$ meson. Hence, the spectra presented here
tend to exclude the hypothesis that the enhancement is due to the
meson-proton interaction. \\
The absolute values of the cross section
for the $pp \to pp\eta^{\prime}$ reaction determined as a function of $s_{pp}$ and $s_{p\eta^{\prime}}$
are given in Tables \ref{tab_pp} and \ref{tab_peta}, and are shown in Figure \ref{fig:norm_pp_peta}.
The determined distributions  differ significantly from the  predictions based on the
homogeneous phase space population~(thick solid line).
Also, the results of calculations including the FSI$_{pp}$ (dotted line)
do not describe
the data underestimating  the cross sections
at large values of $s_{pp}$ and low values of $s_{p\eta}$.
Similarly to the case of $\eta$ meson, a better agreement with the experimental data can be achieved when taking into account
contributions from higher partial waves.
The calculations depicted as solid lines result from a combined analysis (based on the effective Lagrangian approach)
of the production of $\eta$ and $\eta^{\prime}$ mesons
in photo- and hadro-induced reactions \cite{kanzo, kanzo_c11, kanzo1, kanzo2}.
Since the calculation is done in the plane-wave basis, not only the
$^{3}P_{0} \to ^{1}\!\!S_{0}s$ but
also the $^{1}S_{0} \to ^{3}\!\!P_{0}s$ (and all the higher partial-waves) transition\footnote{The transition
  between angular momentum combinations of the initial and final states are
  described according to the conventional notation \cite{meyer} in the
  following way:
\begin{equation}
 ^{2S^{i}+1}L^{i}_{J} \to ^{2S+1}L_{J},l
\label{spectro}
\end{equation}
where, superscript ``i'' indicates the initial state quantities. $S$ denotes
the total spin of nucleons, and $J$ stands for the overall angular momentum of
the system. $L$ and $l$ denote the relative angular momentum of nucleon-nucleon
pair and of the meson relative to the $NN$ system, respectively.}
contributes at the excess energy of 16.4 MeV.\\
On the other hand, one can explain the enhancement seen in the distributions
by an energy dependent production amplitude~\cite{deloff}.
The result indicated by the dashed lines was obtained allowing for a linear energy dependence
of the $^{3}P_{0} \to ^{1}\!\!S_{0}s$ partial wave amplitude and neglecting
other partial waves transitions \cite{deloff}.
Also, Ceci et al.~\cite{ceci} have shown recently that the discussed enhancement in the invariant mass
spectra can be well described by the energy dependence of the production amplitude when
the negative interference between the $\pi$ and the $\eta$ meson exchange amplitudes is assumed.

Within the statistical and systematic error 
bars both model of Nakayama et~al.\cite{kanzo}
and of Deloff~\cite{deloff} describe
the data well although they differ slightly in the predicted shapes.
Taking into account statistical uncertainties only,
one obtains $\chi^{2}$ = 2.1 and $\chi^{2}$ = 4.7
for the comparison of the dashed line and solid line to the data, respectively.
This indicates that perhaps, not only higher partial waves but also the energy dependence of the production amplitude
should be taken into account. 

\section{Summary}
\label{sec:summary}
Using the COSY-11 detector setup and the proton beam of the
cooler synchrotron COSY 
the proton-proton and proton-$\eta^{\prime}$ invariant mass
distributions have been determined for the $pp \to pp\eta^{\prime}$ 
reaction at an excess energy of Q = 16.4 MeV.

Similar to the earlier observation for the $\eta$ meson production the measured
differential cross section distributions ($s_{pp}$ and $s_{p\eta^{\prime}}$) for the $pp \to pp\eta^{\prime}$ reaction
at Q = 16.4 MeV strongly deviate from the predictions based on a
homogeneous population of events over the allowed phase space. Also, the inclusion of the proton-proton final state interaction
is not sufficient to explain the enhancement seen in the range of large $s_{pp}$ values.

Within the achieved uncertainties, the shape of the proton-proton and proton-meson
invariant mass distributions determined for the $\eta^{\prime}$ meson is essentially the same to that 
established previously for the $\eta$ meson.
Since the enhancement is similar in both cases, and the strength of proton-$\eta$ and proton-$\eta^{\prime}$ interaction
is different \cite{swave, pk_phd}, one can conclude that the observed enhancement
is not caused by a proton-meson interaction.

Finally, calculations assuming a significant contribution of P-wave in the
final state~\cite{kanzo},
and models including 
energy dependence of the production amplitude~\cite{deloff,ceci}, reproduce the data
within error bars equally well.
Therefore, on the basis of the presented invariant mass distributions, it is not possible
to disentangle univocally which of the discussed models is more appropriate.
As pointed out in \cite{kanzo}, future measurements of the spin correlation coefficients 
should help disentangle these two model results in a model independent way.\\

{\bf Acknowledgements:}\\
The work was partially supported by the
European Co\-mmu\-nity-Research Infrastructure Activity
under the FP6 and FP7 programmes (Hadron Physics,
RII3-CT-2004-506078, PrimeNet No. 227431.), by
the Polish Ministry of Science and Higher Education under grants
No. 3240/H03/2006/31  and 1202/DFG/2007/03,
by the German Research Foundation (DFG),
by the FFE grants from the Research Center J{\"u}lich,
and by the virtual institute "Spin and strong QCD" (VH-VI-231).

\begin{table}
\begin{center}
\begin{tabular}{c l}
\hline
  $s_{pp}~[GeV^{2}/c^{4}]$  & $\frac{d\sigma}{ds_{pp}}~[\mu $b/GeV$^{2}$/c$^{4}]$\\
\hline
3.5215 & 1.01 $\pm~$ 0.11$_{stat}$ $\pm~$ 0.12$_{sys}$\\
3.5245 & 4.28 $\pm~$ 0.25$_{stat}$ $\pm~$ 0.56$_{sys}$\\
3.5275 & 4.06 $\pm~$ 0.26$_{stat}$ $\pm~$ 0.59$_{sys}$\\
3.5305 & 3.78 $\pm~$ 0.26$_{stat}$ $\pm~$ 0.55$_{sys}$\\
3.5335 & 3.52 $\pm~$ 0.26$_{stat}$ $\pm~$ 0.52$_{sys}$\\
3.5365 & 2.74 $\pm~$ 0.25$_{stat}$ $\pm~$ 0.46$_{sys}$\\
3.5395 & 1.99 $\pm~$ 0.23$_{stat}$ $\pm~$ 0.40$_{sys}$\\
3.5425 & 2.75 $\pm~$ 0.25$_{stat}$ $\pm~$ 0.40$_{sys}$\\
3.5455 & 2.23 $\pm~$ 0.23$_{stat}$ $\pm~$ 0.37$_{sys}$\\
3.5485 & 2.66 $\pm~$ 0.25$_{stat}$ $\pm~$ 0.39$_{sys}$\\
3.5515 & 1.96 $\pm~$ 0.21$_{stat}$ $\pm~$ 0.30$_{sys}$\\
3.5545 & 2.18 $\pm~$ 0.23$_{stat}$ $\pm~$ 0.34$_{sys}$\\
3.5575 & 1.93 $\pm~$ 0.22$_{stat}$ $\pm~$ 0.31$_{sys}$\\
3.5605 & 1.62 $\pm~$ 0.22$_{stat}$ $\pm~$ 0.33$_{sys}$\\
3.5635 & 1.76 $\pm~$ 0.20$_{stat}$ $\pm~$ 0.26$_{sys}$\\
3.5665 & 1.66 $\pm~$ 0.19$_{stat}$ $\pm~$ 0.24$_{sys}$\\
3.5695 & 1.76 $\pm~$ 0.19$_{stat}$ $\pm~$ 0.26$_{sys}$\\
3.5725 & 1.39 $\pm~$ 0.16$_{stat}$ $\pm~$ 0.22$_{sys}$\\
3.5755 & 1.12 $\pm~$ 0.14$_{stat}$ $\pm~$ 0.19$_{sys}$\\
3.5785 & 1.07 $\pm~$ 0.11$_{stat}$ $\pm~$ 0.11$_{sys}$\\
3.5815 & 0.72 $\pm~$ 0.09$_{stat}$ $\pm~$ 0.09$_{sys}$\\
3.5845 & 0.013 $\pm~$ 0.004$_{stat}$ $\pm~$ 0.002$_{sys}$\\
\hline
\end{tabular}
\caption{Differential cross section as a function of the squared invariant mass
of the proton-proton system, for the $pp \to pp\eta^{\prime}$ reaction at Q = 16.4 MeV. }
\label{tab_pp}
\end{center}
\end{table}

\begin{table}
\begin{center}
\begin{tabular}{c l}
\hline
  $s_{p\eta^{\prime}}~[GeV^{2}/c^{4}]$  & $\frac{d\sigma}{ds_{p\eta^{\prime}}}~[\mu $b/GeV$^{2}$/c$^{4}]$\\
\hline
3.5945 & 0.14 $\pm~$ 0.02$_{stat}$ $\pm~$ 0.02$_{sys}$\\
3.5975 & 0.76 $\pm~$ 0.06$_{stat}$ $\pm~$ 0.09$_{sys}$\\
3.6005 & 1.25 $\pm~$ 0.09$_{stat}$ $\pm~$ 0.14$_{sys}$\\
3.6035 & 1.32 $\pm~$ 0.10$_{stat}$ $\pm~$ 0.17$_{sys}$\\
3.6065 & 1.62 $\pm~$ 0.12$_{stat}$ $\pm~$ 0.24$_{sys}$\\
3.6095 & 1.28 $\pm~$ 0.13$_{stat}$ $\pm~$ 0.28$_{sys}$\\
3.6125 & 1.69 $\pm~$ 0.15$_{stat}$ $\pm~$ 0.33$_{sys}$\\
3.6155 & 1.91 $\pm~$ 0.16$_{stat}$ $\pm~$ 0.37$_{sys}$\\
3.6185 & 1.87 $\pm~$ 0.15$_{stat}$ $\pm~$ 0.36$_{sys}$\\
3.6215 & 1.94 $\pm~$ 0.15$_{stat}$ $\pm~$ 0.38$_{sys}$\\
3.6245 & 2.62 $\pm~$ 0.18$_{stat}$ $\pm~$ 0.44$_{sys}$\\
3.6275 & 2.33 $\pm~$ 0.18$_{stat}$ $\pm~$ 0.48$_{sys}$\\
3.6305 & 3.37 $\pm~$ 0.19$_{stat}$ $\pm~$ 0.51$_{sys}$\\
3.6335 & 4.96 $\pm~$ 0.22$_{stat}$ $\pm~$ 0.54$_{sys}$\\
3.6365 & 3.93 $\pm~$ 0.21$_{stat}$ $\pm~$ 0.45$_{sys}$\\
3.6395 & 2.21 $\pm~$ 0.17$_{stat}$ $\pm~$ 0.40$_{sys}$\\
3.6425 & 2.93 $\pm~$ 0.16$_{stat}$ $\pm~$ 0.32$_{sys}$\\
3.6455 & 3.34 $\pm~$ 0.18$_{stat}$ $\pm~$ 0.48$_{sys}$\\
3.6485 & 2.96 $\pm~$ 0.15$_{stat}$ $\pm~$ 0.38$_{sys}$\\
3.6515 & 2.30 $\pm~$ 0.13$_{stat}$ $\pm~$ 0.35$_{sys}$\\
3.6545 & 1.27 $\pm~$ 0.08$_{stat}$ $\pm~$ 0.22$_{sys}$\\
3.6575 & 0.22 $\pm~$ 0.03$_{stat}$ $\pm~$ 0.04$_{sys}$\\
\hline
\end{tabular}
\caption{Differential cross section as a function of the squared invariant mass of
the proton-$\eta^{\prime}$ system, for the $pp \to pp\eta^{\prime}$ reaction at Q = 16.4 MeV. }
\label{tab_peta}
\end{center}
\end{table}


\begin{thebibliography}{00}

\bibitem{pdg}
C. Amsler et al., Phys. Lett. {\bf B~667} (2008) 1.
\bibitem{jossop}
C. P. Jessop et al., Phys. Rev. {\bf D~58} (1998) 052002.
\bibitem{branden}
G. Brandenburg et al., Phys. Rev. Lett. {\bf 75} (1995) 3804.
\bibitem{prc69}
P. Moskal et al., Phys. Rev. {\bf C~69} (2004) 025203.
\bibitem{wycech}
A. M. Green, S. Wycech, Phys. Rev. {\bf C~71} (2005) 014001.
\bibitem{swave}
P. Moskal et al., Phys. Lett. {\bf B~482} (2000) 356.
\bibitem{tof41}
M. Abdel-Bary et al., Eur. Phys. J. {\bf A~16} (2003) 127.
\bibitem{fix}
A. Fix, H. Arenh{\"o}vel, Phys. Rev. {\bf C~69} (2004) 014001.
\bibitem{fix2}
A. Fix, H. Arenh{\"o}vel, Nucl. Phys. {\bf A~697} (2002) 277.
\bibitem{kanzo}
K. Nakayama et al., Phys. Rev. {\bf C~68} (2003) 045201.
\bibitem{deloff}
A. Deloff, {Phys. Rev. \bf C~69} (2004) 035206.
\bibitem{ceci}
S. Ceci, A. {\v{S}}varc, B. Zauner,
Acta Phys. Pol. {\bf B} Suppl.  {\bf 2} (2009) 157.
\bibitem{marcin}
M. Zieli{\'n}ski, Diploma Thesis,
e-print arXiv: hep-ex/0807.0576.
\bibitem{ring1}
R. Maier, Nucl. Instr. $\&$ Meth. {\bf A~390} (1997) 1.
\bibitem{cooling}
D. Prasuhn et al., Nucl. Instr. $\&$ Meth. {\bf A~441} (2000) 167.
\bibitem{cooling1}
H. Stockhorst et al., IKP Annual Report (1997) 160.
\bibitem{cooling2}
H. Stockhorst, Shriften des FZ-J{\"u}lich, Matter and Materials {\bf 11} (2002) 176.
\bibitem{pm80}
P. Moskal et al., Phys. Rev. Lett. {\bf 80} (1998) 3202.
\bibitem{b474}
P. Moskal et al., Phys. Lett. {\bf B~474} (2000) 416.
\bibitem{khoukaz}
A. Khoukaz at al., Eur. Phys. J. {\bf A~20} (2004) 345.
\bibitem{brauksiepe}
S. Brauksiepe et al., Nucl. Instr. $\&$ Meth. {\bf A~376} (1996) 397.
\bibitem{dombrowski}
H. Dombrowski et al.,  Nucl. Instr. $\&$ Meth. {\bf A~386} (1997) 228.
\bibitem{klajac11}
P. Klaja et al., AIP Conf. Proc. {\bf 796} (2005) 160.
\bibitem{target1}
A. Khoukaz et al., Eur. Phys. J. {\bf D~5} (1999) 275.
\bibitem{nim}
P. Moskal et al., Nucl. Instr. $\&$ Meth. {\bf A~466} (2001) 448.
\bibitem{pk_phd}
P. Klaja, PhD thesis, e-Print: arXiv:0907.1491.
\bibitem{jpg}
P. Moskal et al., J. Phys. {\bf G~32} (2006) 629.
\bibitem{kanzo_c11}
K. Nakayama, Y. Oh, H. Haberzettl, Acta Phys. Pol. {\bf B} Suppl. {\bf 2} (2009) 23.
\bibitem{kanzo1}
K. Nakayama, H. Haberzettl, Phys. Rev. {\bf C~69} (2004) 065212.
\bibitem{kanzo2}
K. Nakayama, Y. Oh, H. Haberzettl, e-print: arXiv: hep-ph/0803.3169.
\bibitem{meyer}
H. O. Meyer et al., Phys. Rev. {\bf C~63} (2001) 064002.

\end{thebibliography}
\end{document}